\documentclass[journal,11pt,draftcls,onecolumn]{IEEEtran}
\usepackage[font=small]{caption}
\usepackage{blindtext}
\usepackage{algorithm}
\usepackage{algorithmic}
\usepackage{subfig}
\usepackage{amssymb}
\usepackage[pdftex]{graphicx}
\graphicspath{{./}{figures/}}
\usepackage[table]{xcolor}
\usepackage{amsmath}
\usepackage{amssymb}
\usepackage{url}
\usepackage{setspace}
\DeclareMathOperator*{\argmin}{arg\,min}

\def\Pr{\textrm{Pr}}
\def\eg{\emph{e.g. }}
\def\ie{\emph{i.e. }}

\newcommand{\im}[1]{\mathbf{#1}}

\begin{document}
\title{A New Randomness Evaluation Method with Applications to Image Shuffling and Encryption}
\author{Yue~Wu,~\IEEEmembership{Member,~IEEE,}
        {Sos~Agaian},~\IEEEmembership{Senior Member,~IEEE,}\\
        and Joseph~P.~Noonan,~\IEEEmembership{Life Member,~IEEE,}
\thanks{Yue Wu and Joseph P. Noonan is with the Department of Electrical and Computer Engineering, Tufts University, Medford, MA 02155, United States; e-mail: ywu03@ece.tufts.edu. Sos Agaian is with the Electrical and Computer Engineering, University of Texas at San Antonio, San Antonio, TX 78249, United States.}
}

\maketitle

\begin{abstract}
This letter discusses the problem of testing the degree of randomness within an image, particularly for a shuffled or encrypted image. Its key contributions are: 1) a mathematical model
of perfectly shuffled images; 2) the derivation of the theoretical distribution of pixel differences; 3)
a new $Z$-test based approach to differentiate whether or not a test image is perfectly shuffled; and 4) a
randomized algorithm to unbiasedly evaluate the degree of randomness within a given image. Simulation
results show that the proposed method is robust and effective in evaluating the degree of randomness
within an image, and may often be more suitable for image applications than commonly used testing schemes designed for binary data like NIST 800-22. The developed method may be also useful as a first step in determining
whether or not a shuffling or encryption scheme is suitable for a particular cryptographic application.
\end{abstract}
\section{Introduction}
Currently digital images are a major information resource in real life. A large amount of digital images are private, sensitive, or even classified. Much research has been done to ensure security during image transmission or storage. Two typical protection methods are image shuffling (IS) \cite{RPM} and image encryption (IE) \cite{3DCat,AES}, where
IS only scrambles pixel positions, while IE changes both pixel positions and values.
Regardless of the technical details of a protection method, a key question that need to
be answered is `Is this method secure?'. Unfortunately, this raises a dilemma: if a method is secure, then it must be invulnerable to all attacks; however, it is impossible to list all existing
but unknown attacks, or those will be developed in the future. Thus, all we can do is to test whether it is secure under our known attacks.

Among various attacks, one family is statistical attacks \cite{Stinson}, which take advantage of leaked information from poorly protected images. This leaked information always exists, unless a protected image is indistinguishable from a random one. Therefore, the indistinguishability from a random image is a necessary (but not sufficient) condition for a secure protection method, and evaluating the degree of randomness becomes a prerequisite. For example, FIPS 140-2 tests \cite{FIPS140_2} and NIST 800-22 test suite \cite{NIST800_22} are suggested for cryptographic applications. However, they are merely designed for one-dimensional (1D) data encryption (DE), quite different from two-dimensional (2D) IS or IE. For instance, DE expects random 0s and 1s that are independent identically distributed (i.i.d.) with equal probabilities. IS leads to a pixel distribution identical to that of the original image, and thus can be arbitrary. IE should lead to a uniform distribution, but it is unclear that how to tailor pixels from a 2D image into a 1D bit sequence, so that DE evaluation tools can be properly applied. Thus, additional randomness evaluation tools for images need to be developed.

Although some evaluation methods (EM) have been presented, the following four serious challenges still should be considered (summarized from \cite{measure1}): 1) pixel values and positions should be both considered; 2) the randomness degree should effectively reflect the relationship between a shuffled image and the corresponding method; 3) the evaluation should be dependent only on a test image; and 4) the method complexity should be independent of a test image. However, so far an effective approach does not exist \cite{4724672}. Existing EMs for IS and IE can be classified into two groups: subjective methods and objective methods. The former group commonly uses empirical knowledge to quantify the randomness of a shuffled or encrypted image \cite{yu2006new,measure1,measure2}, while the latter group tests the image randomness with respect to derived statistics \cite{NPCR,wu2012local,5648257}. The use of pixel differences for image randomness has been explored for a long time \cite{yu2006new,measure1,measure2}, but these heuristic, subjective methods allow only quantitative evaluations. Nevertheless, all these mentioned methods are either for IS or IE, but not both of them.

In this letter, we propose a new quality EM for both IS and IE using pixel differences, which solves all four mentioned challenges. The rest of the letter is organized as: Sec. II studies the perfectly shuffled image model; Sec. III constructs hypothesis tests for our randomness EM; Sec. IV discusses a special case for IE; Sec. V provides simulation results; and a summary of the paper with conclusions is given in Sec. VI.

\section{Perfectly Shuffled Images}
\subsection{What Is a Perfectly Shuffled Image?}
Before developing a randomness EM, one has to answer the question `\textit{what is a perfectly shuffled image?}'. Altough there are other answers, we believe it is the one with the properties that 1) all neighbor pixels are independent and 2) only random-like patterns exist within the image. This means that pixels in a perfectly shuffled image are i.i.d. just as a random image. Thus, a perfectly shuffled image can be defined as:

{\noindent{\bf Definition 1}}: (A Perfectly Shuffled Image) \textit{An $L$-intensity scale image $\im{X}\!=\!\{x_l\}_{l\in\Omega}$ on a 2D domain $\Omega$ with a pixel distribution ${\cal{P}}\!\!=\!\![p_0\!,p_1\!,\cdots\!,p_{L-1}]$ ($\sum_{k\!=\!0}^{L\!-\!1}p_k\!=\!1$) is called perfectly shuffled, if each pixel is i.i.d. with distribution ${\cal{P}}$, where $p_k$ is the probability of seeing a pixel at the $k^{\rm th}$ scale in $\im{X}$ and $\delta(.)$ is the Kronecker delta function with 1 at 0 and 0 elsewhere.
}
\begin{equation}\label{eq.pk}
    p_k = \textstyle\sum_{l\in\Omega} \delta(x_l-k)/|\Omega|
\end{equation}
\subsection{Theoretical Statistics}
\noindent{\bf{Theorem 1}}. If image $\im{X}$ is perfectly shuffled, then the pixel difference $\rho_{lk}\!\!=\!\!|\!x_l\!-\!x_k\!|$ between two arbitrary distinctive pixels $x_l$ and $x_k$ with $l,k\!\in\!\Omega$ and $l\!\neq\!k$ satisfy the distribution
\begin{equation}\label{eq.rhodist}
    P_d=\Pr(\rho_{lk}\!=\!d)\!=\!\left\{\!\!\!\!\begin{array}{rl} \textstyle\sum_{k = 0}^{L-1}p_k^2\!\!&\!\!\!\!,\textrm{\,if\,} d = 0\\\textstyle2\sum_{k = d}^{L-1}p_{k-d}p_k\!\!&\!\!\!\!,\textrm{\,if\,} d\!\in\![1,L\!-\!1] \end{array}\right..
\end{equation}
\noindent{\bf{Proof}}: Firstly, compute the probability for $d=0$
\begin{eqnarray*}
  \textstyle\Pr(\rho_{lk} = 0)\!\!\!\!\!\!&\!=\!&\!\!\!\!\! \Pr(x_l=x_k) = \textstyle\sum_{k = 0}^{L-1} \Pr(x_l=x_k|x_k) \Pr(x_k)\\
  \!&\overset{i.i.d.}{=}&\! \textstyle\sum_{k = 0}^{L-1}p_kp_k = \textstyle\sum_{k = 0}^{L-1}p_k^2.
\end{eqnarray*}
Secondly, for arbitrary difference $d\neq 0$, we have
\begin{eqnarray*}
  \Pr(\rho_{lk} = d)\!\!\!\!\!\!&\!=\!&\!\!\!\!\!\Pr(x_l=x_k+d)+\Pr(x_l=x_k-d) \\
  \!\!\!\!\!\!&\!=\!&\!\!\!\!\! \textstyle\sum_{k = 0}^{L\!-\!1\!-\!d} \Pr(x_l\!=\!x_k\!+\!d|x_k)\Pr(x_k)\\\!\!\!\!\!\!&\!\!&\!\!\!\!\!+\textstyle\sum_{k = d}^{L\!-\!1} \Pr(x_l\!=\!x_k\!-\!d|x_k)\Pr(x_k)\\
  \!\! \!&\overset{i.i.d.}{=}&\!\!\!\textstyle\sum_{k\!=\!0}^{L\!-\!1\!-\!d} p_{k+d}p_k+\sum_{k\!=\!d}^{L\!-\!1}p_{k\!-\!d}p_k\\
  \!\!\!\!\!\!&\!=\!&\!\!\!\!\! 2\textstyle\sum_{k\!=\!d}^{L\!-\!1}p_{k\!-\!d}p_k
\end{eqnarray*}
Finally, it is noticeable that
\begin{eqnarray*}
  \textstyle\sum_{d=0}^{L-1}\Pr(\rho_{lk} = d)\!\!\!&\!=\!&\!\!\!\!\! \textstyle\sum_{k = 0}^{L-1}p_k^2+2\sum_{d=1}^{L-1}\sum_{k = d}^{L-1}p_{k-d}p_k\\
  \!\!\!\!\!\!&\!=\!&\!\!\!\!\! (\textstyle\sum_{k = 0}^{L-1}p_k)^2 = 1^2 = 1
\end{eqnarray*}
is indeed a discrete probability distribution. \hfill{$\blacksquare$}

\noindent{\textit{Remark}.} 1) The distribution \eqref{eq.rhodist} is image dependent, because ${\cal{P}}$ is image dependent. 2) ${\cal{P}}$ can be obtained from a shuffled image directly, because ${\cal{P}}$ doesnot change before and after IS.

\noindent{\bf{Theorem 2}}. Let $\overline{\rho_m} = \textstyle\sum_{i=1}^{m} \rho_{l_i,k_i}/m$ be the average of the differences for $m$ pairs pixels with $m\!\geq\!30$ and $\cup_{i=1}^m l_{i}\cap \cup_{i=1}^m k_{i}=\emptyset$. Then $\overline{\rho_m}$ follows the Normal distribution
\begin{equation}\label{eq.norm}
   \overline{\rho_m}\sim {\cal{N}}(\mu,\sigma^2/{m})
\end{equation}
with the mean and variance as
\begin{equation}\label{eq.stat}
\textstyle    \mu = \sum_{d=0}^{L-1} dP_d \textrm{\, and \,} \sigma^2 = \sum_{d=0}^{L-1} d^2P_d-\mu^2.
\end{equation}
\noindent{\bf{Proof}}: Straight-forward by using the central limit theorem for i.i.d. $\rho_{l_ik_i}$s following distribution \eqref{eq.rhodist}.\hfill{$\blacksquare$}
\section{Evaluating Image Shuffling Randomness using Pixel Differences}
\subsection{Hypothesis Tests for Perfectly Shuffled Images}
Knowing the theoretical distribution of $\overline{\rho_m}$, we are able to construct hypothesis tests for IS. In particular, we can test image randomness by comparing the theoretical average with the sample average pixel difference of $m$ pairs pixels as follows:
\begin{description}
  \item[$H_0$:] Null hypothesis that image $\im{Y}$ is indistinguishable from a perfectly shuffled image $\im{X}$, \ie $\widetilde{\rho_m} = \overline{\rho_m}$.
  \item[$H_1$:] Alternative hypothesis that  $\widetilde{\rho_m}\neq \overline{\rho_m}$.
\end{description}
where $\widetilde{\rho_m} = \sum_{i=1}^{m}|y_{l_i}-y_{k_i}|/m$ and $\cup_{i=1}^m l_{i}\cap \cup_{i=1}^m k_{i}=\emptyset$.

Since $\overline{\rho_m}\!\sim\!{\cal{N}}(\mu,\sigma^2/{m})$, this hypothesis test is a $Z$-test, where the test statistic can be constructed as
\begin{equation}\label{eq.zstat}
    z = \dfrac{\widetilde{\rho_m}-\mu}{\sigma/\sqrt{m}} \textrm{\,and\,} z\sim{\cal{N}}(0,1).
\end{equation}
Therefore, given a significance level $\alpha$, we can compute the critical values under the null hypothesis as
\begin{equation}\label{eq.criticalval}
    \widetilde{\rho_m}^{*-} = \mu+\Phi_{\alpha/2}^{-1}\sigma/\sqrt{m}\textrm{\,and\,} \widetilde{\rho_m}^{*+} = \mu-\Phi_{\alpha/2}^{-1}\sigma/\sqrt{m}
\end{equation}
where $\Phi^{-1}_{.}$ is the inverse cumulative distribution function of the standard Gaussian ${\cal{N}}(0,1)$. If the sample average $\widetilde{\rho_m}\notin [\widetilde{\rho_m}^{*-},\widetilde{\rho_m}^{*+}]$, we reject the null hypothesis and claim $\im{Y}$ is distinguishable from a perfectly shuffled image.
\subsection{A New Image Shuffling Quality Evaluation Method}
It is noticeable that the proposed hypothesis tests are valid for $m$ pairs of pixels regardless of their spatial configuration, \eg horizontal neighbor pixels, adjacent image blocks etc. In practice, however, we have to unbiasedly test pixels with all possible spatial configurations. Otherwise, we cannot fairly quantify the degree of randomness. One solution is to use the randomized algorithm 1.

\begin{algorithm}
\scriptsize
\begin{algorithmic}[1]
\REQUIRE $\im{Y}$ is a $L$ intensity scale image and $\alpha$ is the significance level
\REQUIRE $N$=\# of tests, $m$ =\# pixel pairs and $T$= \# of evaluation times
\ENSURE $r$ is the success rate for pixels in $\im{Y}$ passing the hypothesis test.
\STATE Find sample intensity distribution ${\cal{P}}\!=\!\{p_k\}_{k\in [0,L-1]}$ for image $\im{Y}$ using \eqref{eq.pk}
\STATE Compute the pixel difference distribution $\Pr(\rho)\!=\!\{P_d\}_{d\in [0,L-1]}$ using \eqref{eq.rhodist}
\STATE Compute the mean $\mu$ and variance $\sigma^2$ using \eqref{eq.stat}
\STATE Compute critical values $\widetilde{\rho_m}^{*-}$ and $\widetilde{\rho_m}^{*+}$ using \eqref{eq.criticalval}
\FOR{$t=1$ to $T$}
\STATE Initialize $r_t = 0$
\FOR{$i = 1$ to $N$}
\STATE Pull $m$ pairs of image pixels with a random configuration w$\backslash$o repetitions \\$\im{Yb}_l\!\!\!=\!\!\!\{y_{l_1},\cdots,y_{l_m}\}$ and $\im{Yb}_k\!=\!\{y_{k_1},\cdots,y_{k_m}\}$.
\STATE Compute $\widetilde{\rho_m}\!=\!\sum_{i=1}^{m}|y_{l_i}-y_{k_i}|/m$
\IF{$\widetilde{\rho_m}\in [\widetilde{\rho_m}^{*-},\widetilde{\rho_m}^{*+}]$}
\STATE $r_t$++
\ENDIF
\ENDFOR
\ENDFOR
\STATE $r =\max_{t\in[1,T]}\{r_t\}/N$.
\end{algorithmic}
\caption{\textbf{$(\alpha,N,m,T)$ Pixel Difference Test for IS}}
\end{algorithm}
Algorithm 1 considers both pixel values and positions, reflects the randomness within a test image, relies only on a test image and has a constant complexity $O(NTm)$. Therefore, it successfully solves all four challenges specified in \cite{measure1}.

Assume $\mathbf{Y}$ is perfectly shuffled. The four parameters in Algorithm 1 have the following impacts. Parameter $\alpha$ is close related to the pass times $r_t$, whose mean and variance are
\begin{equation}\label{eq.rt}
    {\rm{E}}[r_t]=(1\!-\!\alpha)N {\rm{\,and\,}} {\rm{E}}[\!(r_t\!-\!{\rm{E}}[r_t])^2\!]\!=\!N\!\alpha\!(\!1\!-\!\alpha\!).
\end{equation}
Parameter $N$ influences the number of spatial configurations and also the number of pixels used in the evaluation. Thus, a larger $N$ implies a more fair evaluation.

Parameter $m$ has two related effects: 1) the larger $m$ is, the smaller the variance $\sigma^2/m$ is in \eqref{eq.norm}; 2) the larger the $m$ is, the poorer the localization of the test. The former effect prefers a large $m$ to attain a small variance, while the latter one prefers a small $m$ to test a local pixel configuration. Therefore, an optimal $m^*$ should be the one balancing both effects. To do so, we define the loss function \eqref{eq.engm} for $m$, where term $\sigma^2/m$ reflects the loss in the test accuracy, term $m^2/|\Omega|$ reflects the loss in localization capacity, and $\lambda$ is the coefficient indicating the relative importance of the two types of loss.
\begin{equation}\label{eq.engm}
    \Psi(m) = \sigma^2/m+\lambda m^2/|\Omega|
\end{equation}
Hence, the optimal $m^*$ minimizes the loss function $\Psi(m)$
\begin{equation}\label{}
    \textstyle m^* = \argmin_{m\in[1,|\Omega|]} \Psi(m)
\end{equation}
and is of the form \eqref{eq.mstar}, where $\lceil\cdot\rceil$ is the integer rounding function towards infinity.
\begin{equation}\label{eq.mstar}
    m^* = \lceil\sqrt[3]{\sigma^2|\Omega|/(2\lambda)}\rceil
\end{equation}

Finally, parameter $T$ plays an important role to reduce the type I error (false positive error) of the evaluation. As $T\!\rightarrow\!\infty$, the type I error goes to zero.
\begin{eqnarray}\label{eq.T}
  \!\!\!&\!\!\!\!\!\lim\limits_{T\!\rightarrow\!\infty}\Pr(r\!<\!1\!-\!\alpha|H_0)\!=\! \lim\limits_{T\!\rightarrow\!\infty}\Pr(\max_{t\in[1,T]}\{r_t\}\!<{\rm E}[r_t]|H_0)=\!\!\textstyle\prod_{t\!=\!1}^{T\!\!\rightarrow\! \infty}\!\Pr(r_t\!<\!{\rm E}[r_t]|H_0)\!=\!0
\end{eqnarray}

\section{Special Case for Image Encryption}
Because ideally an encrypted image is random-like with a uniform pixel distribution \cite{NPCR,wu2012local}, the derived statistics, hypothesis tests, and the proposed EM are also applicable to IE, by simply using $\cal{P}$=${\cal{U}}[0,L-1]$ (\ie $p_k = 1/L$ in \eqref{eq.pk}).

Specifically, the pixel difference distribution for perfectly encrypted images follows the triangular distribution \eqref{eq.rhodiste} and the corresponding average pixel difference for $m$ pair pixels $\overline{\rho_m^e}$ follows the Normal distribution $\overline{\rho_m^e}\!\sim\!{\cal N}(\mu_e\!,\sigma^2_e\!/\!m)$.
\begin{equation}\label{eq.rhodiste}
    \Pr^e(\rho_{lk}\!=\!d)\!=\!\left\{\!\!\!\!\begin{array}{rl} \textstyle 1/L \!\!&\!\!\!\!,\textrm{\,if\,} d = 0\\\textstyle2(L-d)/L^2\!\!&\!\!\!\!,\textrm{\,if\,} d\!\in\![1,L\!-\!1] \end{array}\right.
\end{equation}
\begin{equation}\label{eq.stat2}
\textstyle    \mu_e\! =\! {(L^2\!-\!1)/(3L)}\textrm{\,and\,} \sigma^2_e \!= \!{(L^2\!-\!1)(L^2\!+\!2)/(18L^2)}
\end{equation}

Table \ref{tab.1} gives the theoretical numerical results of $\mu_e$ and $\sigma_e$ under various image types, where $L=$2, 256 and 65536 refer to the binary image,  8-bit grayscale image and  16-bit grayscale image, respectively. The critical values of the hypothesis tests can be computed as
\begin{equation}\label{eq.criticalval2}
    \widetilde{\rho_m^e}^{*-} = \mu_e+\Phi_{\alpha/2}^{-1}{\sigma_e\over\sqrt{m_e^*}}\textrm{\,and\,} \widetilde{\rho_m^e}^{*+} = \mu_e-\Phi_{\alpha/2}^{-1}{\sigma_e\over\sqrt{m_e^*}}
\end{equation}
where $m_e^*$ is the optimal number of independent pixel pairs used in the quality measure and can be computed via \eqref{eq.mstar}.

\begin{table}[h]
\caption{Pixel Differences Statistics in Perfectly Encrypted Images}\label{tab.1}
\centering
\scriptsize
\begin{tabular}{r|ccc}
  \hline
  &\multicolumn{3}{c}{\bf Image Intensity Levels ${L}$}\\\hline
  \bf{Stat.} & \bf{2} &\bf{256} &\bf{65536}\\\hline
  $\mu_e$ & 0.500 & 85.332 & 21845.333\\
  $\sigma_e$ & 0.500 & 60.340 & 15446.983\\
  \hline
\end{tabular}
\end{table}
\section{Simulation Results}
We apply the proposed randomness EM (Algorithm 1) for IS and IE to real data. All following experiments are done under the MATLAB r2012a environment. Test images are square size 8-bit grayscale (\ie $L$=256) from the USC-SIPI \textit{miscellaneous} dataset (detail descriptions are available online \footnotemark[1]).
\footnotetext[1]{is available under the page \url{http://sipi.usc.edu/database/database.php?volume=misc} as of 10/22/2012.}

The parameters $N$, $\alpha$ and $T$ in Algorithm 1 are set to 1000, 0.05, and 10 respectively. Parameter $m$ is optimized with respect to each test image with $\lambda=\mu/L$ for IS and $\lambda=\mu_e/L$ for IE. Related statistics for performing the quality evaluation are given in Table \ref{tab.stat}, where the critical values are computed with respect to $\alpha$=.05.

Under these parameter settings, it is noticeable that $$\Pr(r_t<{\rm E}[r_t]|H_0)\!=\!\textstyle\sum_{s=0}^{949}{1000\choose s} {.95}^s{.05}^{1000\!-\!s}\!=\!.4625$$
and thus $$\textstyle\Pr(r\!<\!1\!-\!\alpha|H_0)={.4625}^{10}\!<\!5\!/\!10000$$
indicating a very small type I error of the evaluation method.

\begin{table}[h]
\caption{Statistics of Pixel Differences for Test Images}\label{tab.stat}
\scriptsize\centering
\begin{tabular}{@{}r@{}m{.05cm}@{}r@{}m{.05cm}@{}|@{}m{.05cm}@{}r@{}m{.05cm}@{}r@{}m{.05cm}@{}r@{}m{.05cm}@{}r@{$\sim$}r@{}m{.05cm}@{}|@{}m{.05cm}@{}r@{}m{.05cm}@{}r@{}m{.05cm}@{}r@{}m{.05cm}@{}r@{$\sim$}r}
  \hline\hline
\multicolumn{3}{c}{\bf{Image Info.}}&&&\multicolumn{8}{c}{\bf Stat. for IS}&&&\multicolumn{8}{c}{\bf Stat. for IE}\\
\multicolumn{1}{c}{\bf{Name}}&&\bf{Side}&&&$\mu$&&$\sigma$&&$m^*$&&\multicolumn{2}{c}{${\widetilde{\rho_{m^*}}^{*\pm}}$}&&&$\mu_e$&&$\sigma_e$&&$m^*_e$&&\multicolumn{2}{c}{${\widetilde{\rho_{m^*_e}}^{*\pm}}$}\\\hline
\bf{5.1.09}&&256&&&29.58&&25.76&&574&&27.47&31.69&&&85.33&&60.34&&711&&80.90&89.77\\
\bf{5.1.10}&&256&&&51.54&&38.15&&619&&48.53&54.54&&&85.33&&60.34&&711&&80.90&89.77\\
\bf{5.1.11}&&256&&&33.82&&32.31&&638&&31.31&36.33&&&85.33&&60.34&&711&&80.90&89.77\\
\bf{5.1.12}&&256&&&58.51&&55.96&&766&&54.54&62.47&&&85.33&&60.34&&711&&80.90&89.77\\
\bf{5.1.13}&&256&&&51.24&&94.08&&1132&&45.76&56.72&&&85.33&&60.34&&711&&80.90&89.77\\
\bf{5.1.14}&&256&&&47.60&&36.37&&616&&44.73&50.47&&&85.33&&60.34&&711&&80.90&89.77\\
\bf{5.2.08}&&512&&&43.75&&35.99&&998&&41.52&45.98&&&85.33&&60.34&&1128&&81.81&88.85\\
\bf{5.2.09}&&512&&&40.58&&38.27&&1066&&38.29&42.88&&&85.33&&60.34&&1128&&81.81&88.85\\
\bf{5.2.10}&&512&&&61.91&&46.46&&1054&&59.11&64.72&&&85.33&&60.34&&1128&&81.81&88.85\\
\bf{5.3.01}&&1024&&&66.54&&47.79&&1664&&64.24&68.84&&&85.33&&60.34&&1790&&82.54&88.13\\
\bf{5.3.02}&&1024&&&37.88&&32.88&&1565&&36.25&39.51&&&85.33&&60.34&&1790&&82.54&88.13\\
\bf{7.1.01}&&512&&&29.48&&24.41&&879&&27.87&31.10&&&85.33&&60.34&&1128&&81.81&88.85\\
\bf{7.1.02}&&512&&&17.00&&26.26&&1109&&15.46&18.55&&&85.33&&60.34&&1128&&81.81&88.85\\
\bf{7.1.03}&&512&&&27.98&&26.10&&935&&26.31&29.65&&&85.33&&60.34&&1128&&81.81&88.85\\
\bf{7.1.04}&&512&&&36.38&&33.02&&1002&&34.34&38.43&&&85.33&&60.34&&1128&&81.81&88.85\\
\bf{7.1.05}&&512&&&39.71&&29.46&&902&&37.79&41.64&&&85.33&&60.34&&1128&&81.81&88.85\\
\bf{7.1.06}&&512&&&37.93&&27.97&&885&&36.09&39.77&&&85.33&&60.34&&1128&&81.81&88.85\\
\bf{7.1.07}&&512&&&26.19&&21.92&&851&&24.72&27.66&&&85.33&&60.34&&1128&&81.81&88.85\\
\bf{7.1.08}&&512&&&21.85&&26.18&&1018&&20.24&23.45&&&85.33&&60.34&&1128&&81.81&88.85\\
\bf{7.1.09}&&512&&&40.67&&30.50&&916&&38.70&42.65&&&85.33&&60.34&&1128&&81.81&88.85\\
\bf{7.1.10}&&512&&&29.12&&24.38&&882&&27.51&30.73&&&85.33&&60.34&&1128&&81.81&88.85\\
\bf{7.2.01}&&1024&&&21.27&&29.33&&1758&&19.89&22.64&&&85.33&&60.34&&1790&&82.54&88.13\\
\bf{boat.512}&&512&&&49.85&&43.27&&1081&&47.27&52.43&&&85.33&&60.34&&1128&&81.81&88.85\\
\bf{elaine.512}&&512&&&52.65&&38.35&&979&&50.25&55.05&&&85.33&&60.34&&1128&&81.81&88.85\\
\bf{gray21.512}&&512&&&88.72&&62.95&&1145&&85.07&92.36&&&85.33&&60.34&&1128&&81.81&88.85\\
\bf{numbers.512}&&512&&&70.28&&51.91&&1088&&67.19&73.36&&&85.33&&60.34&&1128&&81.81&88.85\\
\bf{ruler.512}&&512&&&49.94&&101.20&&1902&&45.40&54.49&&&85.33&&60.34&&1128&&81.81&88.85\\
\bf{testpat.1k}&&1024&&&85.28&&64.98&&1881&&82.35&88.22&&&85.33&&60.34&&1790&&82.54&88.13\\
  \hline\hline
\end{tabular}
\end{table}

\begin{table}[h]
\caption{Quality Evaluation Scores (${10}^{-3}$) for Tested Methods}\label{tab.stat2}
\scriptsize\centering
\begin{tabular}{@{}r@{}m{.1cm}@{}|@{}m{.1cm}@{}c@{}m{.1cm}@{}|@{}m{.1cm}@{}c@{}m{.1cm}@{}c@{}m{.1cm}@{}c@{}m{.1cm}@{}c@{}m{.1cm}@{}|@{}m{.1cm}@{}c@{}m{.1cm}@{}c@{}m{.1cm}@{}c@{}}\hline
&&&{\bf Original}&&\multicolumn{9}{c|}{\bf IS}&\multicolumn{6}{c}{\bf IE}\\
{\bf{Test Image}}&&&&&&\bf{RPM}&&\bf{RCS}&&\bf{ATM}&&\bf{SS}&&&\bf{CBC}&&\bf{ECB}&&\bf{LME}\\
\hline
\bf{5.1.09}&& &\underline{479}& & &960&&\underline{904}&&\underline{489}&&959& & &960&&964&&959\\
\bf{5.1.10}&&&\underline{894}&&&957&&\underline{937}&&\underline{918}&&959&&&961&&960&&960\\
\bf{5.1.11}&&&\underline{223}&&&960&&\underline{835}&&\underline{210}&&960&&&957&&957&&956\\
\bf{5.1.12}&&&\underline{271}&&&961&&\underline{836}&&\underline{272}&&962&&&957&&961&&956\\
\bf{5.1.13}&&&\underline{871}&&&961&&\underline{898}&&\underline{908}&&957&&&960&&\underline{489}&&955\\
\bf{5.1.14}&&&\underline{695}&&&960&&\underline{905}&&\underline{720}&&960&&&958&&955&&959\\
\bf{5.2.08}&&&\underline{668}&&&958&&\underline{836}&&\underline{673}&&960&&&964&&964&&965\\
\bf{5.2.09}&&&\underline{877}&&&960&&952&&\underline{928}&&959&&&965&&952&&956\\
\bf{5.2.10}&&&\underline{340}&&&955&&\underline{889}&&\underline{526}&&962&&&964&&963&&950\\
\bf{5.3.01}&&&\underline{387}&&&957&&\underline{937}&&\underline{418}&&958&&&960&&962&&957\\
\bf{5.3.02}&&&\underline{603}&&&958&&\underline{943}&&\underline{649}&&961&&&959&&957&&962\\
\bf{7.1.01}&&&\underline{472}&&&958&&\underline{934}&&\underline{503}&&958&&&961&&965&&961\\
\bf{7.1.02}&&&\underline{354}&&&968&&\underline{936}&&\underline{354}&&965&&&961&&963&&966\\
\bf{7.1.03}&&&\underline{433}&&&957&&\underline{910}&&\underline{470}&&966&&&957&&961&&964\\
\bf{7.1.04}&&&\underline{460}&&&960&&\underline{915}&&\underline{509}&&958&&&964&&964&&964\\
\bf{7.1.05}&&&\underline{759}&&&959&&\underline{920}&&\underline{789}&&957&&&956&&962&&959\\
\bf{7.1.06}&&&\underline{681}&&&961&&\underline{937}&&\underline{715}&&961&&&959&&958&&956\\
\bf{7.1.07}&&&\underline{804}&&&955&&\underline{939}&&\underline{844}&&961&&&971&&969&&969\\
\bf{7.1.08}&&&\underline{560}&&&959&&\underline{896}&&\underline{525}&&964&&&956&&951&&966\\
\bf{7.1.09}&&&\underline{116}&&&963&&\underline{864}&&\underline{113}&&961&&&960&&956&&961\\
\bf{7.1.10}&&&\underline{650}&&&957&&\underline{913}&&\underline{662}&&959&&&962&&960&&961\\
\bf{7.2.01}&&&\underline{598}&&&958&&\underline{929}&&\underline{668}&&963&&&961&&954&&953\\
\bf{boat.512}&&&\underline{302}&&&957&&\underline{843}&&\underline{319}&&958&&&956&&960&&957\\
\bf{elaine.512}&&&\underline{499}&&&963&&\underline{936}&&\underline{594}&&961&&&959&&956&&956\\
\bf{gray21.512}&&&\underline{62}&&&956&&\underline{852}&&\underline{71}&&955&&&960&&\underline{567}&&970\\
\bf{numbers.512}&&&\underline{707}&&&961&&\underline{916}&&\underline{777}&&958&&&965&&967&&962\\
\bf{ruler.512}&&&\underline{531}&&&962&&\underline{801}&&\underline{701}&&958&&&959&&\underline{299}&&961\\
\bf{testpat.1k}&&&\underline{222}&&&957&&\underline{894}&&\underline{269}&&959&&&957&&\underline{628}&&956\\
\hline
\bf{Mean}&&&518.5&&&959.2&&900.3&&556.9&&960.0&&&960.3&&893.7&&959.9\\
\bf{Standard Deviation}&&&229.8&&&2.8&&40.9&&239.5&&2.5&&&3.4&&172.2&&4.8\\
\hline
\end{tabular}
\end{table}

Tested IS methods are the random permutation method (RPM) \cite{RPM}, row and column shuffling (RCS) \cite{ye2010image}, Arnold transform method (ATM) \cite{3DCat}, and Sudoku shuffling (SS) \cite{wuSudoku}. Meanwhile, we also use the IE methods including the logistic map encryption (LME) \cite{wu013014}, the AES cipher \cite{AES} with the electronic codebook (ECB) mode and the cipher-block chaining (CBC) mode, respectively. It is worthy to point out that the RPM and the AES-CBC are classic methods for IS and IE, respectively. Therefore, it is not surprising to see good performance scores for these two methods. In contrast, the RCS and ATM are known to have mesh-like and slant patterns, respectively. Thus we shall expect to see low scores for these two methods. The AES-ECB method is known to be insecure for patterned images, because it cannot encrypt two identical blocks to different random patterns with the same key.

\begin{figure}[!h]
  \scriptsize\centering
  \begin{tabular}{c@{}m{.05cm}@{}c@{}m{.05cm}@{}c@{}m{.05cm}@{}c}
  \includegraphics[width=3.75cm]{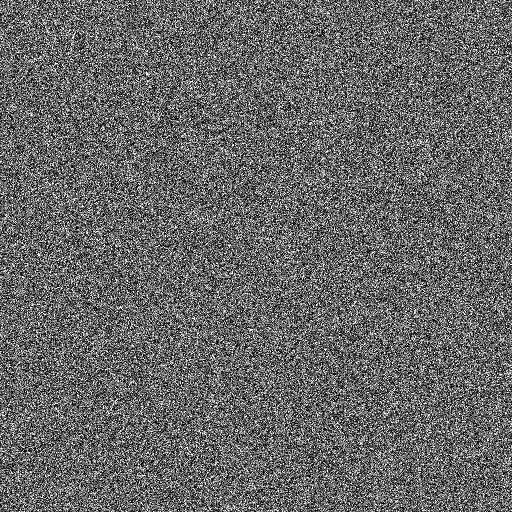}&&  \includegraphics[width=3.75cm]{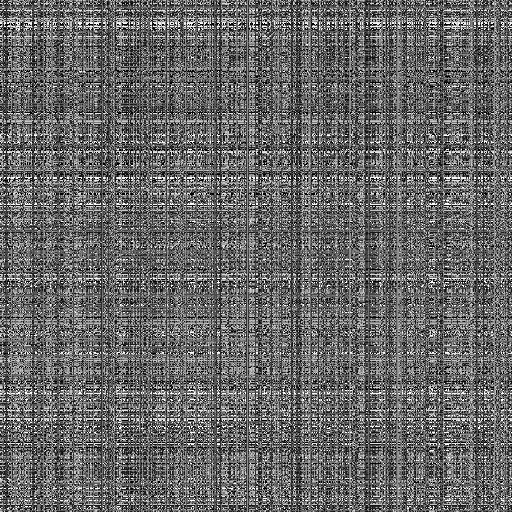} &&  \includegraphics[width=3.75cm]{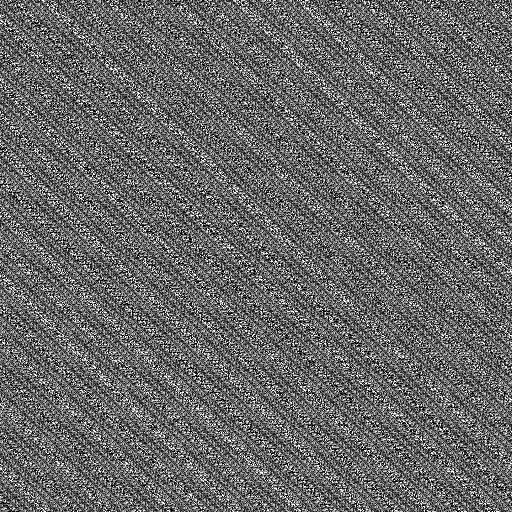}&&\includegraphics[width=3.75cm]{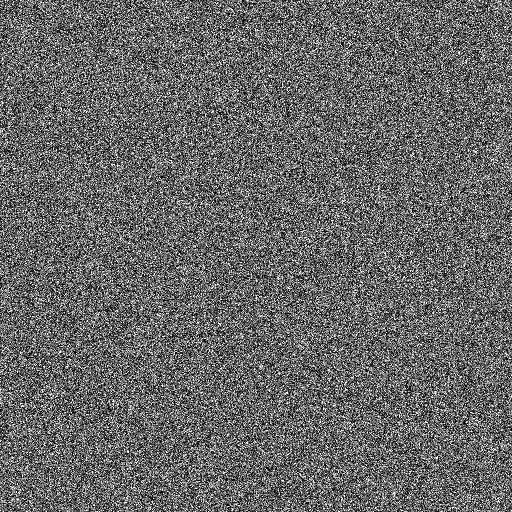}\\
  (a1) numbers.RPM && (b1) numbers.RCS &&(c1) numbers.ATM && (d1) numbers.SS\\
  \includegraphics[width=3.75cm]{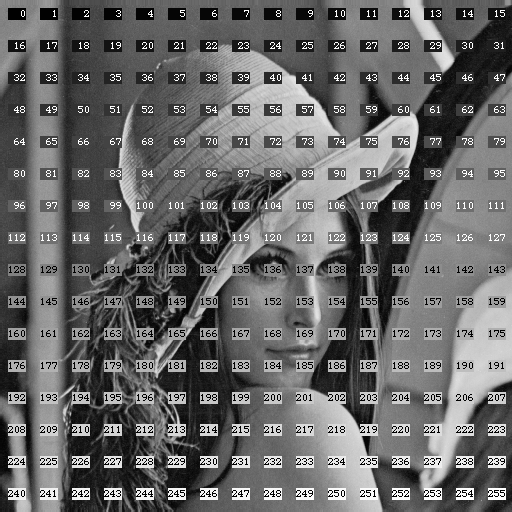}&&\includegraphics[width=3.75cm]{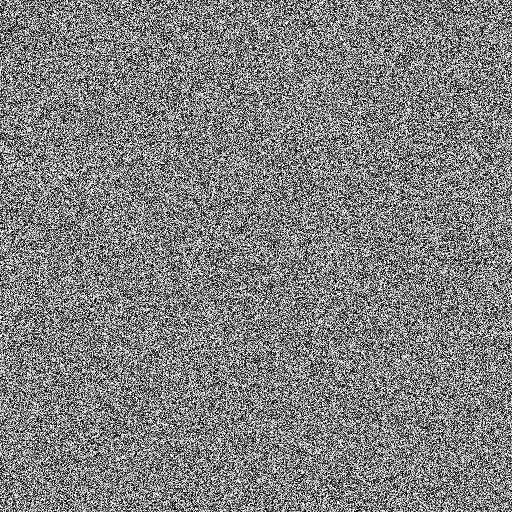}&&\includegraphics[width=3.75cm]{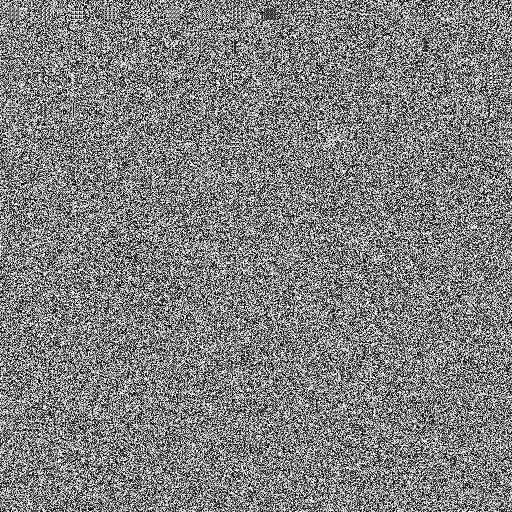}&&\includegraphics[width=3.75cm]{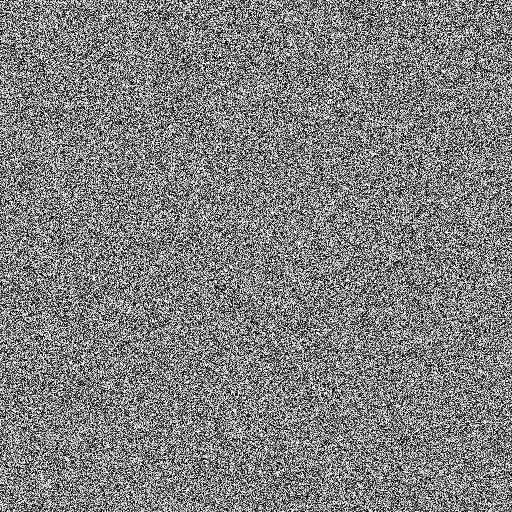}\\
  (e1) Original image: numbers.512 && (f1) numbers.CBC && (g1) numbers.ECB &&(h1) numbers.LME\\
  \includegraphics[width=3.75cm]{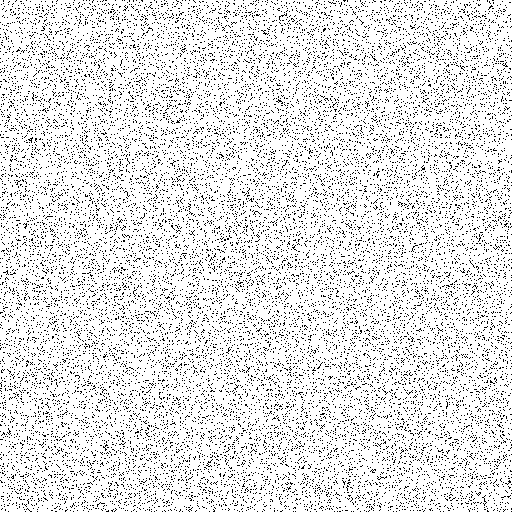}&&  \includegraphics[width=3.75cm]{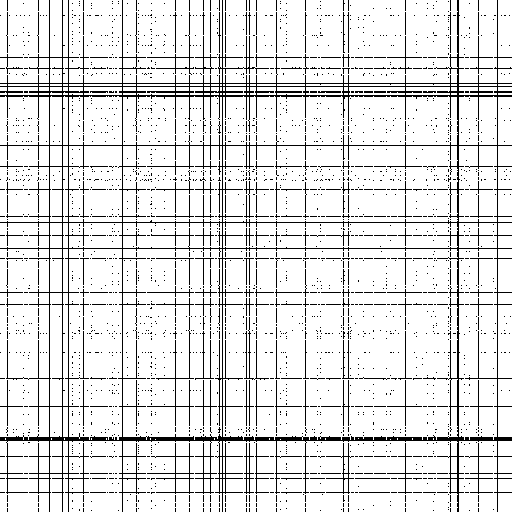} &&  \includegraphics[width=3.75cm]{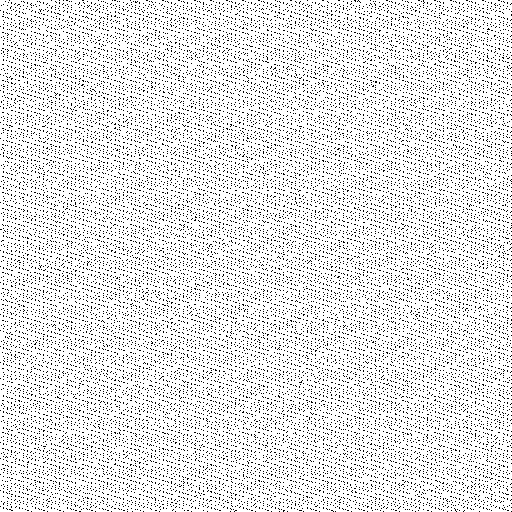}&&\includegraphics[width=3.75cm]{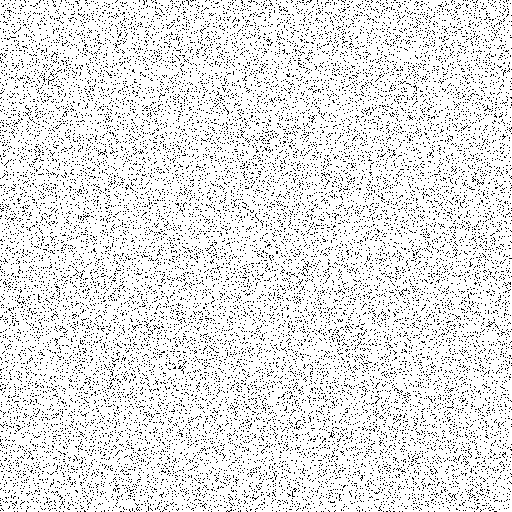}\\
  (a2) ruler.RPM && (b2) ruler.RCS &&(c2) ruler.ATM && (d2) ruler.SS\\
  \includegraphics[width=3.75cm]{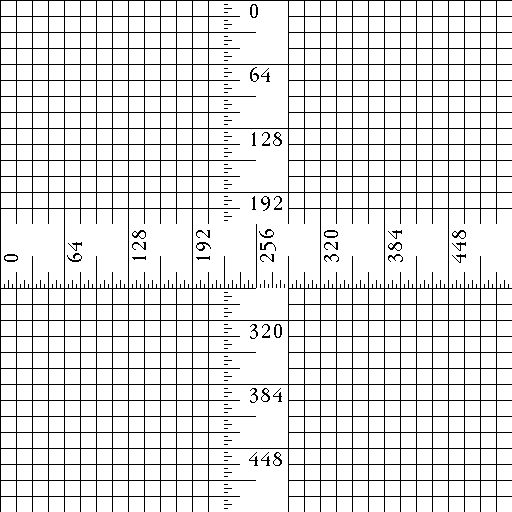}&&\includegraphics[width=3.75cm]{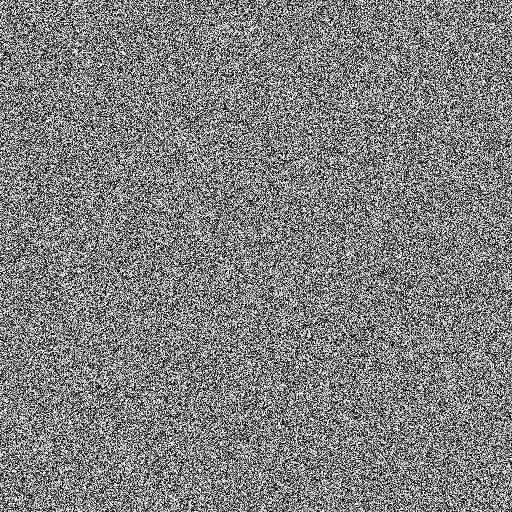}&&\includegraphics[width=3.75cm]{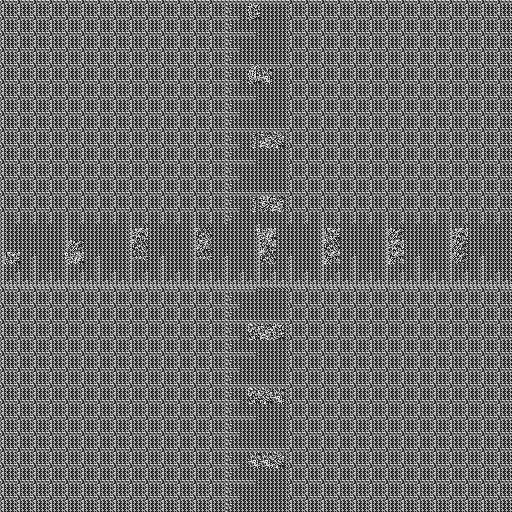}&&\includegraphics[width=3.75cm]{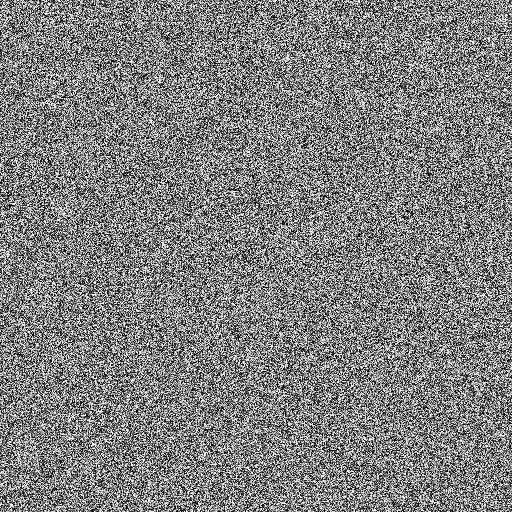}\\
  (e2) Original image: ruler.512 && (f2) ruler.CBC && (g2) ruler.ECB &&(h2) ruler.LME\\
  \end{tabular}
  \caption{Shuffled and encrypted images}\label{fig.result2}
\end{figure}
Evaluation results for these methods are listed in Table III. As one can see, RPM and AES-CBC have their evaluation scores all above 1-$\alpha$=.95. In contrast, none of evaluation scores for the RCS and ATM is above .95 (emphasized by underlines), indicating that one can distinguish their shuffled images from those perfectly shuffled. The scores for these two methods vary greatly for different test images, especially for the ATM, whose IS performance is known to be related to the period of a random Arnold transform. The AES-EBC has good scores for most images but bad scores for highly structured images like \textit{ruler.512} and \textit{testpat.1k}. The SS and LME both have good evaluation scores above .95 indicating that their shuffled or encrypted images are indistinguishable from random-like. Fig. \ref{fig.result2} gives sample results of these methods. Once again, the indistinguishability of shuffled or encrypted images from random-like ones are only a necessary condition for a method to be secure. All these results are sufficient to conclude the methods RCS, ATM and ECB are insecure.

\section{Conclusion}
We introduced a new randomness (quality) evaluation method for IS. It tests whether an image is indistinguishable from a perfectly shuffled one, whose image pixels are i.i.d. everywhere. The proposed method can be used as a prerequisite test to claim an IS or IE method is secure. Simulation results indicate that this method is accurate and effective for quality evaluation for IS and IE methods. Moreover, it can be used as an indicator of quality control. That is, if a shuffled or encrypted image fails to reach the expected evaluation score, it should be re-shuffled or encrypted using another key. Further, the developed method may be useful as a first step in determining whether or not an IE scheme is suitable for a particular cryptographic application. The proposed method is of a constant complexity and can be easily implemented. Thus, it is well suited for practical applications. A MATLAB implementation can be provided on request.
\newpage
\bibliographystyle{IEEEtran}
\bibliography{reference}
\end{document}